\documentstyle[12pt]{article}

\begin{document}
\input{psfig.sty}
\begin{flushright}
\baselineskip=12pt
MADPH-98-1051 \\
\end{flushright}

\begin{center}
\vglue 1.5cm
{\Large\bf Soft Terms in M-theory \\}
\vglue 2.0cm
{\Large  Tianjun Li}
\vglue 1cm
\begin{flushleft}
Department of Physics, University of Wisconsin, Madison, WI 53706
\end{flushleft}
\end{center}

\vglue 1.5cm
\begin{abstract}
We discuss dilaton and Moduli
dominant SUSY breaking scenarios in M-theory. In addition, for
the nonperturbative superpotential from gaugino condensation, 
we discuss the soft terms
in the simplest model (only S and T moduli fields) and in
the $T^6/Z_{12}$ model from M-theory. From the phenomenology
consideration, we suggest massless scalar SUSY breaking scenario.

\end{abstract}

\vspace{0.5cm}
\begin{flushleft}
\baselineskip=12pt
April 1998\\
\end{flushleft}
\newpage
\setcounter{page}{1}
\pagestyle{plain}
\baselineskip=14pt

\section{Introduction}
M-theory on $S^1/Z_2$ suggested by Horava and 
Witten~\cite{HW} seems to be a better candidate than the 
weakly coupled string to explain the low energy physics 
and unification of all the fundamental interactions~\cite{Witten}.
Many recent studies in M-theory compactification and 
its phenomenology
implications~\cite{Horava, AQR, EDCJ, LT, LOW, CKM, NOY,
BD, KC, AQ,VK, EFN, MP, LLN, TIAN, LOD, DVN, BKL, BL, 
BJ, JEDVN}
 suggest that this may be the right 
track for super unification.

In this paper we discuss the soft terms in M-theory derived
models. In supergravity Grand Unified Theory (GUT) models,
there are several soft terms to determine the low energy 
physics: universal gaugino mass $M_{1/2}$, universal scalar mass
$m_0$, trilinear coupling A-term, bilinear coupling B-term(s).
Compared to the spectrum of low energy super particles, the input 
parameters are really very few. However, a futher reduction of
parameters may be realized in string theory. In the weakly coupled
string, there are two simplifying choices: (1) No-scale scenario
~\cite{no-scale, JLDVN} with
 $m_0=A=0$; (2) Dilaton scenario~\cite{ABIM, VSKJL} with
$M_{1/2}=-A=M_0/\sqrt 3$. In our present discussion, we consider 
dilaton and moduli 
dominant SUSY breaking scenarios in M-theory. In 
addition, with the
non-perturbative superpotential from the gaugino condensation
~\cite{LOW, BL}, we obtain
the soft terms: $M_{1/2}$, $M_0$ and $A$ in terms of the
 hidden sector gauge coupling ( at the scale of $M_{GUT}$ ) 
in units of the gravitino mass. We find
no consistent solution for  
moduli dominant SUSY breaking and only one choice  
of dilaton dominant supersymmetry breaking, the latter gives 
\begin{eqnarray}
M_{1/2} \simeq 2 M_0 \simeq - A  ~,~\,
\end{eqnarray}
 which differs
from previously obtained results~\cite{ABIM, VSKJL}. There might exist 
further Moduli or other Dilaton dominant SUSY breaking models, for 
there might exist other trigger of the SUSY breaking
and we do not have a realistic model yet.
Furthermore, we suggest a massless scalar SUSY breaking scenario.
Because the gauge coupling in the hidden sector can not be
too small, this scenario does not have flavour changing neutral current
(FCNC) problem because derivations from zero mass for the scalar
masses in different families 
are much smaller
than the gaugino mass~\cite{NOY}. In this scenario, the soft terms are:
\begin{eqnarray} 
A \simeq - 0.77 M_{1/2} ~;~ M_0^i \simeq 0 ~,~\,
\end{eqnarray}
where i=1, 2, 3 for three families. 

\section{Dilaton/Moduli Dominant SUSY Breaking}
The Lagrangian for the Yang-Mills field in M-theory is
~\cite{HW}:
\begin{eqnarray}
L_B&=&
{1\over\displaystyle 2\pi (4\pi \kappa^2)^{2\over 3}}
\int_{M^{10}_i}d^{10}x\sqrt g {1\over 4}F_{AB}^aF^{aAB} ~,~\,
\end{eqnarray}
where $\kappa^2$ denotes the 11-dimensional gravitational coupling
and  i=1,2 for the observable sector, hidden sector.
In the simplest model, the GUT scale coupling is related to 
moduli fields S and T by~\cite{NOY, LOD}:
\begin{eqnarray}
\alpha_{GUT}&=&{1\over\displaystyle {4\pi Re(S + \alpha T)}} ~,~ \,
\end{eqnarray}
The definition of $\alpha$ is~\cite{LOD}:
\begin{eqnarray}
\alpha &=& {\rho \over {16V}} ({\kappa \over {4 \pi}})^{2/3}
\int \omega \wedge tr( R \wedge R ) ~,~\,
\end{eqnarray}
and $\rho$ is the fifth ( or eleventh ) dimension radius,
V is the internal Calabi-Yau volume and $\omega$ is
the K\"ahler form.
 Similarly, the gauge coupling in the hidden sector
at the GUT scale is
\begin{eqnarray}
\alpha_H&=&{1\over\displaystyle {4\pi Re(S - \alpha T)}} ~.~\,
\end{eqnarray}
Inverting the relations in Eq.s (4) and (6), we obtain:
\begin{eqnarray}
ReS &=& {1\over {8 \pi}}({1\over \alpha_{GUT}}+
{1\over \alpha_H}) ~,~ \,
\end{eqnarray}
and
\begin{eqnarray}
ReT &=& {1\over {8 \pi \alpha }}({1\over \alpha_{GUT}}-
{1\over \alpha_H}) ~.~ \,
\end{eqnarray}
The K\"ahler potential $K$, the gauge kinetic function $f$ and 
the superpotential $W$
in the simplest compactification of M-theory are~\cite{NOY, LOD}:
\begin{eqnarray}
K &=& \hat K + \tilde K |C|^2 ~,~ \,
\end{eqnarray}
\begin{eqnarray}
\hat K &=&  -\ln\,[S+\bar S]-3\ln\,[T+\bar T] ~,~\,
\end{eqnarray}
\begin{eqnarray}
\tilde K &=& ({3\over\displaystyle {T+\bar T}} +
{\alpha\over\displaystyle {S+\bar S}}) |C|^2  ~,~ \,
\end{eqnarray}
\begin{eqnarray}
Ref^O_{\alpha \beta} &=& Re(S + \alpha T)\, \delta_{\alpha \beta} ~,~\,
\end{eqnarray}
\begin{eqnarray}
Ref^H_{\alpha \beta} &=& Re(S - \alpha T)\, \delta_{\alpha \beta} ~,~\,
\end{eqnarray}
\begin{eqnarray}
W= d_{x y z} C^x C^y C^z ~,~\,
\end{eqnarray}
where $C^x$ are the matter fields.
We do not specify the exact trigger of the SUSY breaking,
but just assume SUSY breaking by the F terms of the dilaton
and moduli.
Then, we obtain the gaugino mass $M_{1/2}$, scalar mass $M_0$
and the trilinear parameters A by standard soft term
formulae~\cite{ABIM, VSKJL}.
We introduce the following parametrizations of $F^S$ and 
$F^T$:
\begin{eqnarray}
F^S &=& \sqrt 3 ~M_{3/2} ~C ~( S+\bar S) ~sin\theta ~,~\,
\end{eqnarray}
\begin{eqnarray}
F^T &=& M_{3/2} ~C ~( T +\bar T) ~cos\theta ~,~\,
\end{eqnarray}
where C is given in terms of the tree-level vacuum energy density
$V$ given by $C^2=1+{V\over\displaystyle {3 M_{3/2}^2}}$.
After normalizing the observable
fields, the soft terms are given by~\cite{CKM, NOY, LOW}
\footnote{Because the bilinear parameter $B$ depends on the specific
mechanism which could generate the associated $\mu$ term~\cite{CKM},
we do not discuss it here. In fact, if we had such term, we can 
discuss it by the same way in this paper. It is~\cite{BKL}
\begin{eqnarray}
B_{\mu} &=&  M_{3/2} ( - 3 C cos\theta - \sqrt 3 C sin\theta
+{{6 C cos\theta } \over\displaystyle
{ 3 + x }}
\nonumber\\&&
+ {{2 \sqrt 3 C sin\theta x} \over\displaystyle 
{3 + x}} - 1 + F^S {\mu^S \over \mu } + F^T {\mu^T \over \mu} ) ~,~ \,
\end{eqnarray}
where $\mu^S = \partial_S \mu$ and $\mu^T = \partial_T \mu$.
If one consider 
${\mu^S \over \mu } = {\mu^T \over \mu} = 0 $, then one obtain:
\begin{eqnarray}
B_{\mu} &=&  M_{3/2} ( - 3 C cos\theta - \sqrt 3 C sin\theta
+{{6 C cos\theta } \over\displaystyle
{ 3 + x }}
\nonumber\\&&
+ {{2 \sqrt 3 C sin\theta x} \over\displaystyle 
{3 + x}} - 1 ) ~.~ \,
\end{eqnarray}
}:
\begin{eqnarray}
M_{1/2}&=&{{\sqrt 3 C M_{3/2}} \over\displaystyle {1+x}}
(sin\theta +{x\over \sqrt 3} cos\theta ) ~,~\,
\end{eqnarray}
\begin{eqnarray}
M_0^2&=&V + M_{3/2}^2 - 
\nonumber\\&&
{{3 C M_{3/2}^2} \over\displaystyle 
{(3+x)^2}}(x(6+x) sin^2\theta + ( 3 + 2x ) cos^2\theta
- 2 \sqrt 3 x ~sin\theta ~cos\theta) ~,~ \,
\end{eqnarray}
\begin{eqnarray}
A&=&- {{\sqrt 3 C M_{3/2}} \over\displaystyle 
{(3+x)}} ((3-2x) sin\theta + \sqrt 3~ x ~cos\theta) ~,~ \,
\end{eqnarray}
Here, the quantity x is defined as
\begin{eqnarray}
x={{\alpha ( T + \bar T )} \over\displaystyle { S + \bar S }} 
={{\alpha_{GUT}^{-1} \alpha_H - 1}\over\displaystyle 
{\alpha_{GUT}^{-1} \alpha_H + 1}} ~,~\,
\end{eqnarray}
Because $\alpha > 0 $ and the gauge coupling in the hidden
sector can not be infinity or negative, we have
\begin{eqnarray}
1 \geq x \geq 0 ~.~ \,
\end{eqnarray}

Taking the Cosmology constant is zero, i. e., $ V = 0$, we have $C=1$. 
We choose $\alpha_{GUT}=1/25$
for the following discussion.
Generically, the dynamics of the hidden sector may give rise to
both $< F^S >$ and $< F^T > $, but one type of F term often dominates.
 Therefore, we concentrate on the two limiting cases:
dilaton dominant SUSY breaking ($F^T=0$) and the moduli dominant 
SUSY breaking ($F^S=0$)~\cite{ABIM, VSKJL}:

(I) Dilaton dominant SUSY breaking scenario ($F^T=0$). In this 
case, the soft terms relations become
\footnote{
If one consider 
${\mu^S \over \mu } = {\mu^T \over \mu} = 0 $, then one obtain:
\begin{eqnarray}
B_{\mu} &=& - {{3 + 3 \sqrt 3 - (\sqrt 3 - 1) x}
\over\displaystyle { 3 + x }} M_{3/2}  ~.~ \,
\end{eqnarray}
}
\begin{eqnarray}
M_{1/2}&=&{{\sqrt 3  M_{3/2}} \over\displaystyle {1+x}} ~,~\,
\end{eqnarray}
\begin{eqnarray}
M_0^2&=& M_{3/2}^2 - {{ 3  M_{3/2}^2} \over\displaystyle 
{(3+x)^2}} ~x ~(6+x) ~,~  \,
\end{eqnarray}
\begin{eqnarray}
A&=&- {{\sqrt 3  M_{3/2}} \over\displaystyle 
{3+x }} (3-2x) ~.~\,
\end{eqnarray}
The values of $M_{1/2}$, $M_0$ and
$A$ versus $\alpha_H$ are plotted in Fig. 1 in units of 
gravitino mass. When $\alpha_H = \alpha_{GUT}$,
we obtain the predicted relation in the weakly coupled 
string~\cite{ABIM, VSKJL}.
 When the
gauge coupling in the hidden sector increases, $M_{1/2}$,
$M_0$ and $A$ decrease. 
From the requirement $M_0^2 > 0$, we obtain that
$x$ is smaller than $-3+3 \sqrt 6 /2 $ or 0.67423. 
In the limit of $M_0 = 0$, we find that 
$A/M_{1/2} \simeq - 0.77 $, and the $\alpha_H$ 
is about 0.2056. 

(II) Moduli dominant SUSY breaking scenario ( $F^S = 0$ ). In 
this case, the soft terms relations become
\footnote{
If one consider 
${\mu^S \over \mu } = {\mu^T \over \mu} = 0 $, then one obtain:
\begin{eqnarray}
B_{\mu} &=& - {{ 6 + 4 x }
\over\displaystyle { 3 + x }} M_{3/2}  ~.~ \,
\end{eqnarray}
}
\begin{eqnarray}
M_{1/2}&=&{x\over\displaystyle {1+x}} M_{3/2} ~,~ \,
\end{eqnarray}
\begin{eqnarray}
M_0&=& {x\over\displaystyle {3+x}} M_{3/2} ~,~\,
\end{eqnarray}
\begin{eqnarray}
A&=&- {{ 3x } \over\displaystyle {3+x}} M_{3/2} ~.~\,
\end{eqnarray}
Therefore, we have
\begin{eqnarray}
 M_0/A=-1/3 ~;~ 3 \geq M_{1/2}/M_0 \geq 2 ~.~\,
\end{eqnarray} 
From Fig. 2, we see that if the coupling in the hidden
sector is strong enough (i. e., $\alpha_H > 0.4$),  $M_{1/2}/M_0$
is about 2. In addition, $M_{1/2}$,
$M_0$ and $A$ increase wth the
gauge coupling in the hidden sector. 

\section{Hidden Sector Gaugino Condensation}
The nonperturbative superpotential in the simplest model was
calculated recently, and its form is~\cite{LOW}:
\begin{eqnarray}
W_{np} &=& h~ exp(-{{8 \pi^2}\over\displaystyle C_2(Q)}
(S- \alpha T)) ~,~\,
\end{eqnarray}
where the gauge group in the hidden sector
is Q ( we do not consider massless Q matter fields 
except gluons and gluinos ). 
In the simplest model Q is $E_8$. $C_2(Q)$ is the
quadratic Casimir of Q and for $E_8$, $C_2 = 30$.

Next, we can calculate $F^S$ and $F^T$ from 
$W_{np}~\cite{ABIM, VSKJL}$. Noticing that
 $< C^a > =0$, we have
\begin{eqnarray}
\bar F^S &=& - e^{\hat K /2} W_{np} (S+\bar S)^2
({{8 \pi^2} \over\displaystyle {C_2(Q)}} +
{1\over\displaystyle {S+\bar S}}) ~,~\,
\end{eqnarray}
\begin{eqnarray}
\bar F^T &=& - e^{\hat K /2} W_{np} (T+\bar T)^2
(-{{8 \pi^2 \alpha } \over\displaystyle {3 C_2(Q)}} +
{1\over\displaystyle { T +\bar T}}) ~.~\,
\end{eqnarray}
Taking the ratio $F^S/F^T$, we obtain the value of $tan\theta$
in terms of $\alpha_H $
\begin{eqnarray}
tan\theta &=& {1\over\displaystyle \sqrt 3} 
{{1+{2\pi \over\displaystyle {C_2(Q)}}(\alpha_{GUT}^{-1}
+\alpha_H^{-1})}\over\displaystyle 
{1-{2\pi \over\displaystyle {3 C_2(Q)}}(\alpha_{GUT}^{-1}
- \alpha_H^{-1})}} ~.~ \,
\end{eqnarray}

We now discuss the soft terms of compactification of M-theory on 
$T^6/Z_{12}$~\cite{BL, BJ}. 
Although this is a toy model, it has three
families, so it may shed light on the realistic model.
The K\"ahler potential is :
\begin{eqnarray}
K &=& \hat K + 
\tilde K_{i\bar i} C_{i a}^* C_i^a ~,~\,
\end{eqnarray}
where i=1, 2, 3 is the family index,
\begin{eqnarray}
\hat K = -\ln\,[S+\bar S ]
-\sum_{i=1}^3 \ln\, [T_i +\bar T_i] ~,~ \,
\end{eqnarray}
and 
\begin{eqnarray}
\tilde K_{i\bar i}=  ( 2 + {2\over 3} {1\over {S+\bar S}} 
(\sum_{j=1}^3 \alpha_j (T_j + \bar T_j)))
{1\over\displaystyle {T_i +\bar T_i}} ~.~\, 
\end{eqnarray}
The gauge kinetic function in this model is
\begin{eqnarray}
Ref^O_{\alpha \beta} &=& Re(S + \alpha_1 T_1 
+ \alpha_2 T_2 + \alpha_3 T_3)\, \delta_{\alpha \beta} ~,~ \,
\end{eqnarray}
\begin{eqnarray}
Ref^H_{\alpha \beta} &=& Re(S - \alpha_1 T_1 
- \alpha_2 T_2 - \alpha_3 T_3)\, \delta_{\alpha \beta} ~,~ \,
\end{eqnarray}
and the superpotential is
\footnote{We use this form of the non-perturbative superpotential,
because the calculation in~\cite{LOW} should be general.}
\begin{eqnarray}
W= d_{x y z} C_1^x C_2^y C_3^z + 
h ~exp(-{{8 \pi}\over\displaystyle {C_2(Q)}} 
(S - \alpha_1 T_1 - \alpha_2 T_2 - \alpha_3 T_3)) ~,~\,
\end{eqnarray}
We parametrize $F^S$ and $F^T$ as follows:
\begin{eqnarray}
F^S &=& {\sqrt 3} sin\theta ~ M_{3/2}
  (S +\bar S)  ~,~\,
\end{eqnarray}  
\begin{eqnarray}
F^{T_1} &=& {\sqrt 3} cos\theta ~ sin\beta_1 ~ 
 M_{3/2}  (T_1 +\bar T_1)  ~,~\,
\end{eqnarray}  
\begin{eqnarray}
F^{T_2} &=& {\sqrt 3} cos\theta ~ cos\beta_1 ~ sin\beta_2 ~
 M_{3/2}  (T_2 +\bar T_2)  ~,~\,
\end{eqnarray}  
\begin{eqnarray}
F^{T_3} &=& {\sqrt 3} cos\theta ~ cos\beta_1 ~ cos\beta_2 ~
 M_{3/2}  (T_3 +\bar T_3)  ~,~\,
\end{eqnarray}  
Then the normalized soft gaugino masses $M_{1/2}$, the 
un-normalized 
soft scalar masses $M_0^{i2}$, and the trilinear parameters
$A_{a b c}$ are given by~\cite{BL}:
\begin{eqnarray}
M_{1/2}&=&{{F^S + \sum_{i=1}^3 \alpha_i F^{T_i}}
\over\displaystyle { (S+ \bar S) +
\sum_{j=1}^3 \alpha_j ( T_j + \bar T_j) }} ~,~\,
\end{eqnarray} 
\begin{eqnarray}
M_0^{i2}&=&(m_{3/2}^2 +V) \tilde K_{\bar i i}
\nonumber\\&& - \bar F^{\bar m} (\partial_{\bar m}
\partial_n \tilde K_{\bar i i}- 
\partial_{\bar m} \tilde K_{\bar i i}
{1\over\displaystyle \tilde K_{\bar i i}}
\partial_n \tilde K_{\bar i i}) F^n ~,~\,
\end{eqnarray}
\begin{eqnarray}
A_{ a b c }= F^m ( \hat K_m - \sum_{i=1}^3 
{1\over \tilde K_{\bar i i}} 
\partial_m \tilde K_{\bar i i}) ~,~\,
\end{eqnarray}
The non-zero $\hat K_m$, $\partial_m \tilde K_{\bar i i} $
and $\partial_{\bar m} \partial_n \tilde K_{\bar i i} $ were 
calculated in Ref.~\cite{BL}.

To discuss the number of variables in this model, we
introduce:
\begin{eqnarray}
x_i&=& {{\alpha_i (T_i + \bar T_i)} 
\over\displaystyle {S+\bar S}} ~,~\,
\end{eqnarray}
where i=1, 2, 3, 
with the condition that
\begin{eqnarray}
x&=& x_1 + x_2 + x_3 ~.~\,
\end{eqnarray}
Thus only two of the $x_i$ are independent
variables when we consider $\alpha_H$ as a fundamental variable.
The angles $\theta$, 
$\beta_1$ and $\beta_2$ can be calculated in a way similar 
to the simplest model.
We can express the 5 soft terms ( $M_{1/2}$, $M_0^i$ ( i = 1, 2, 3 ),
A ) in terms of four variables ( $\alpha_H$, $M_{3/2}$ and
two of the $x_i$ 
( for example: $x_1$ and $x_2$ ) ). The problem with this
general model is that the universality of scalar masses may be violated
and there are too many variables to make predictions.
To obtain universal scalar masses, we assume that:
\begin{eqnarray} 
x_i = x/3 ~;~ i= 1, 2, 3 ~.~\,
\end{eqnarray}
From the result
\begin{eqnarray}
F^{T_i}&=& e^{\hat K/2} W_{np} ( T_i + \bar T_i )^2
({{8 \pi^2 \alpha_i}\over\displaystyle {C_2(Q)}}
-{1\over\displaystyle {T_i +\bar T_i}}) ~.~\,
\end{eqnarray}
the ratios $F^{T_2}/F^{T_3}$ and $F^{T_1}/F^{T_2}$ lead to
\begin{eqnarray}
tan\beta_2 = 1 ~;~ tan\beta_1 = {\sqrt 2 \over\displaystyle 2} ~.~\,
\end{eqnarray}
Therefore, we can write the $F^{T_i}$ as
\begin{eqnarray}
F^{T_i} &=& cos\theta ~ M_{3/2} (T_i + \bar T_i ) ~.~\,
\end{eqnarray}
After some calculations, we are able to show that the scalar
masses are universal and that the soft terms are given by 
Eqs.(19), (20), (21) and $\theta$ is given by Eq. (36).

Because $x$ is a function of $\alpha_H$, 
$M_{1/2}, M_0$ and $A$ are given by $\alpha_H$ in units of the
gravitino mass. In addition, it is obvious that moduli dominant 
SUSY breaking scenario does not exist for the
numerator in $tan\theta$ is larger than zero. In the simplest
model and the $T^6/Z_{12}$ model, the gauge group in the hidden
sector is $E_8$, for which $C_2(Q)=30$. The result is shown
in the Fig. 3.
In order that the scalar mass square non
negative, $\alpha_H$ can not be larger than 
about 0.18.  There is
only one choice of exact dilaton dominant SUSY breaking,
which is:
\begin{eqnarray}
M_{1/2}=1.236 M_{3/2} ~;~ M_0 = 0.5776 M_{3/2} ~;~  \,
\end{eqnarray}
\begin{eqnarray}
A=-1.119 M_{3/2} ~,~ \alpha_H = 0.09 ~.~ \,
\end{eqnarray}
In order to avoid FCNC problems that might arise from the violation
of the universal scalar masses in three families ( although this
kind of the violation might be very small),  it is reasonable assume
that:
\begin{eqnarray} 
\alpha_1 (T_1 + \bar T_1 ) \simeq \alpha_2 (T_2 + \bar T_2 )
\simeq  \alpha_3 (T_3 + \bar T_3 ) ~.~\,
\end{eqnarray}
If the gaugino mass is very large compared to the scalar masses,
i. e., $ M_{1/2} >> M_0^i $, the universal radiative contribution
to the scalar masses from the gaugino mass by RGE running may wash out the 
non-universal scalar masses and avoid FCNC~\cite{NOY}. In addition,
the gauge coupling in the hidden sector can not be too small
as to go to the weakly coupled case and can not be too large, because
the gaugino condesation scale is~\cite{LOW, NOY}
\begin{eqnarray}
\Lambda ~\sim~ M_{GUT}^3 exp(-{{2 \pi}\over\displaystyle 
{C_2(Q) \alpha_H}}) ~,~ \,
\end{eqnarray}
and the gravitino mass is about
\begin{eqnarray} 
M_{3/2} \sim { \Lambda \over\displaystyle M_{pl}^2} ~.~ \,
\end{eqnarray}
In oder to have the gravitino mass at TeV or hundred GeV range, we can
not  have large $\alpha_H$ or too small $\alpha_H$ for a realistic
model where $C_2(Q)$ might be about 12 as $Q$ is $E_6$, and
when one consider the massless Q matter fields in hidden sector, the
total factor may be smaller.
We propose
the scalar massless scenario ( strictly speaking, very small scalar
masses scenario ) as the most realistic. In this case, the soft 
terms and $\alpha_H$ are given by:
\begin{eqnarray}
M_{1/2}=1.022 M_{3/2}~;~ M_0 = 0.003 M_{3/2}~;~  \,
\end{eqnarray}
\begin{eqnarray}
A=-0.776 M_{3/2}~,~ \alpha_H = 0.1801 ~.~ \,
\end{eqnarray}

In a realistic model, the gauge group in the hidden sector
might not be $E_8$, If consider $E_6$ as the gauge group
in the hidden sector, the quadratic Casimir of $E_6$ is 12. 
In this case that $\alpha_H$ can not be larger than 0.151, and the 
scalar massless scenario is:
\begin{eqnarray}
M_{1/2}=0.989 M_{3/2} ~;~ M_0 = 0.008 M_{3/2}~;~  \,
\end{eqnarray}
\begin{eqnarray}
A=-0.761 M_{3/2}~,~ \alpha_H = 0.1522 ~.~ \,
\end{eqnarray}
In comparison,  the dilaton dominant SUSY breaking in this case is:
\begin{eqnarray}
M_{1/2}= 1.534 M_{3/2}~;~ M_0 = 0.870 M_{3/2} ~;~  \,
\end{eqnarray}
\begin{eqnarray}
A=-1.517 M_{3/2}~,~ \alpha_H = 0.052~.~ \,
\end{eqnarray}

From above discussion, in the 
dilaton dominant SUSY breaking scenario, 
the approximate soft term relations are
\begin{eqnarray}
M_{1/2} \simeq 2 M_0 \simeq - A ~.~\,
\end{eqnarray}
In the massless scalar scenario, the soft term relations are
\begin{eqnarray}
A \simeq - 0.77 M_{1/2} ~;~ M_0^i \simeq 0 ~.~ \,
\end{eqnarray}
It will be interesting to explore the phenomenology consequences
of these relationships which are under investigations.

\section*{Acknowledgments}
I would like to thank V. Barger very much for helpful discussion and comments.
This research was supported in part by the U.S.~Department of Energy under Grant No.~DE-FG02-95ER40896 and in part by the University of Wisconsin Research Committee with funds granted by the Wisconsin Alumni Research Foundation.

\begin{figure}
\centerline{\psfig{file=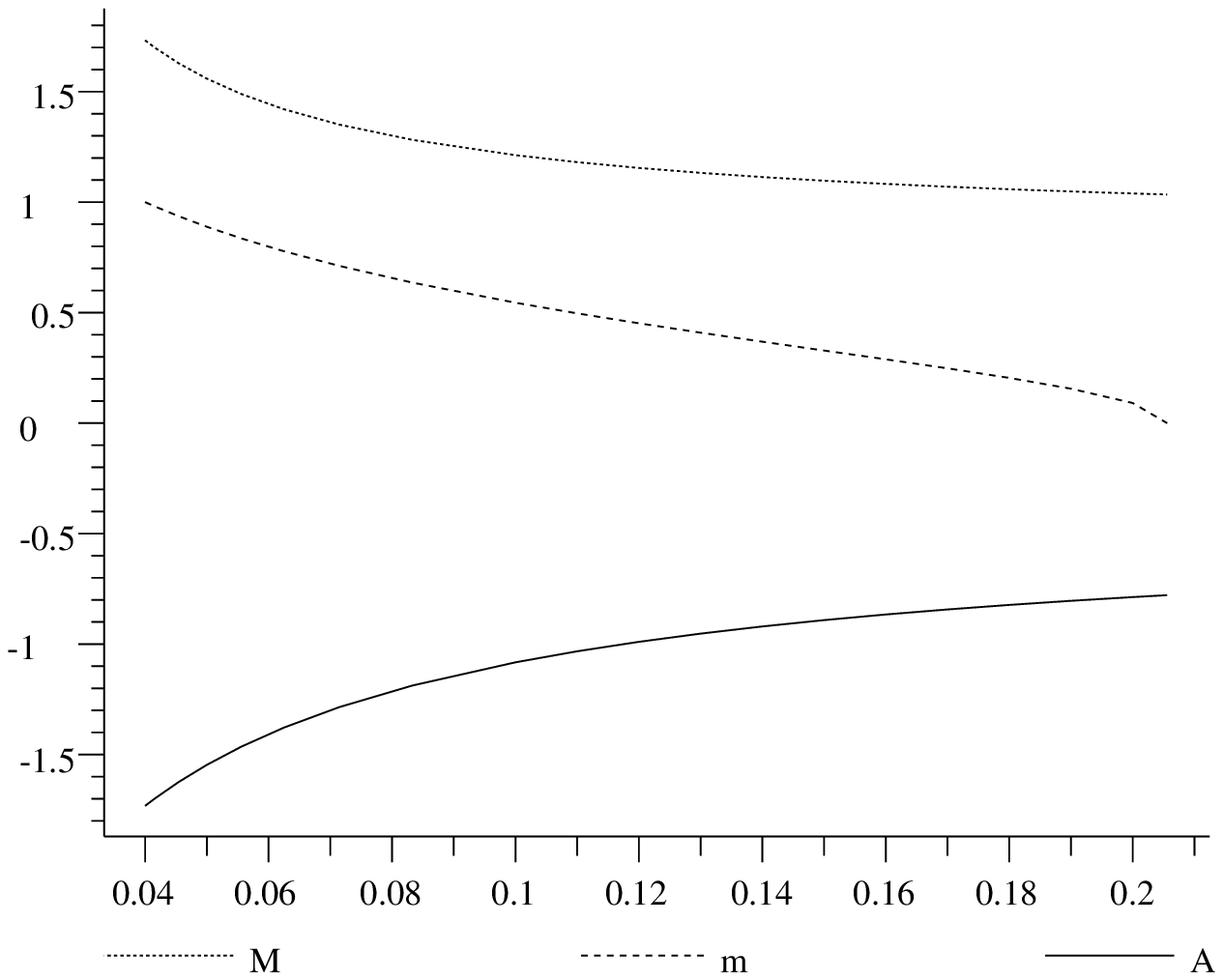,width=15cm}}
\bigskip
\caption[]{The 
gaugino mass $M$, scalar mass $m$, and the trilinear coupling 
$A$ versus $\alpha_H$ in units of the gravitino mass
in the dilaton dominant SUSY Breaking scenario,.}
\label{diagrams}
\end{figure}

\begin{figure}
\centerline{\psfig{file=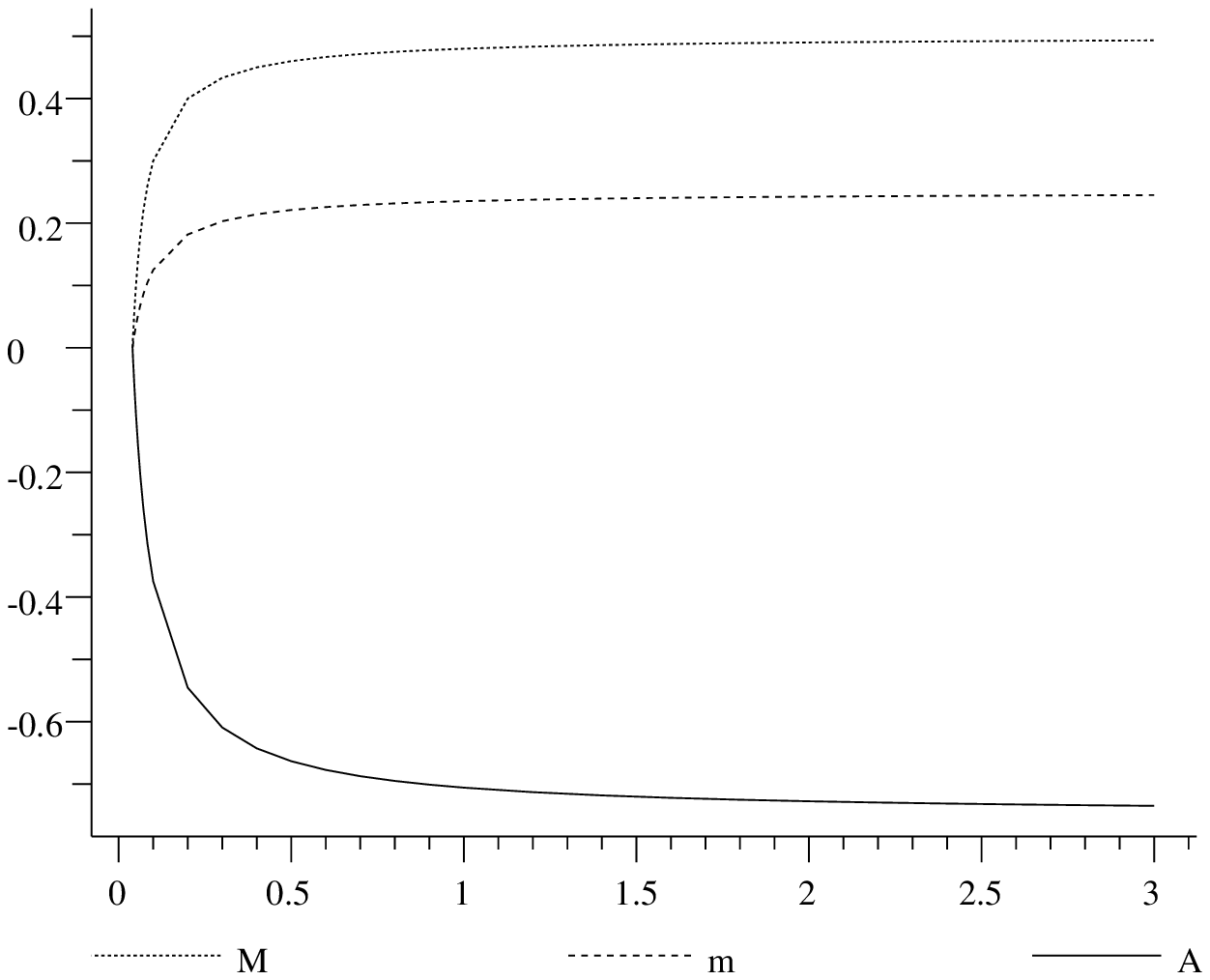,width=15cm}}
\bigskip
\caption[]{ 
The gaugino mass $M$, scalar mass $m$, and the trilinear coupling 
$A$ versus $\alpha_H$ in units of the gravitino mass
in the moduli dominant SUSY Breaking scenario.}
\label{diagrams}
\end{figure}

\begin{figure}
\centerline{\psfig{file=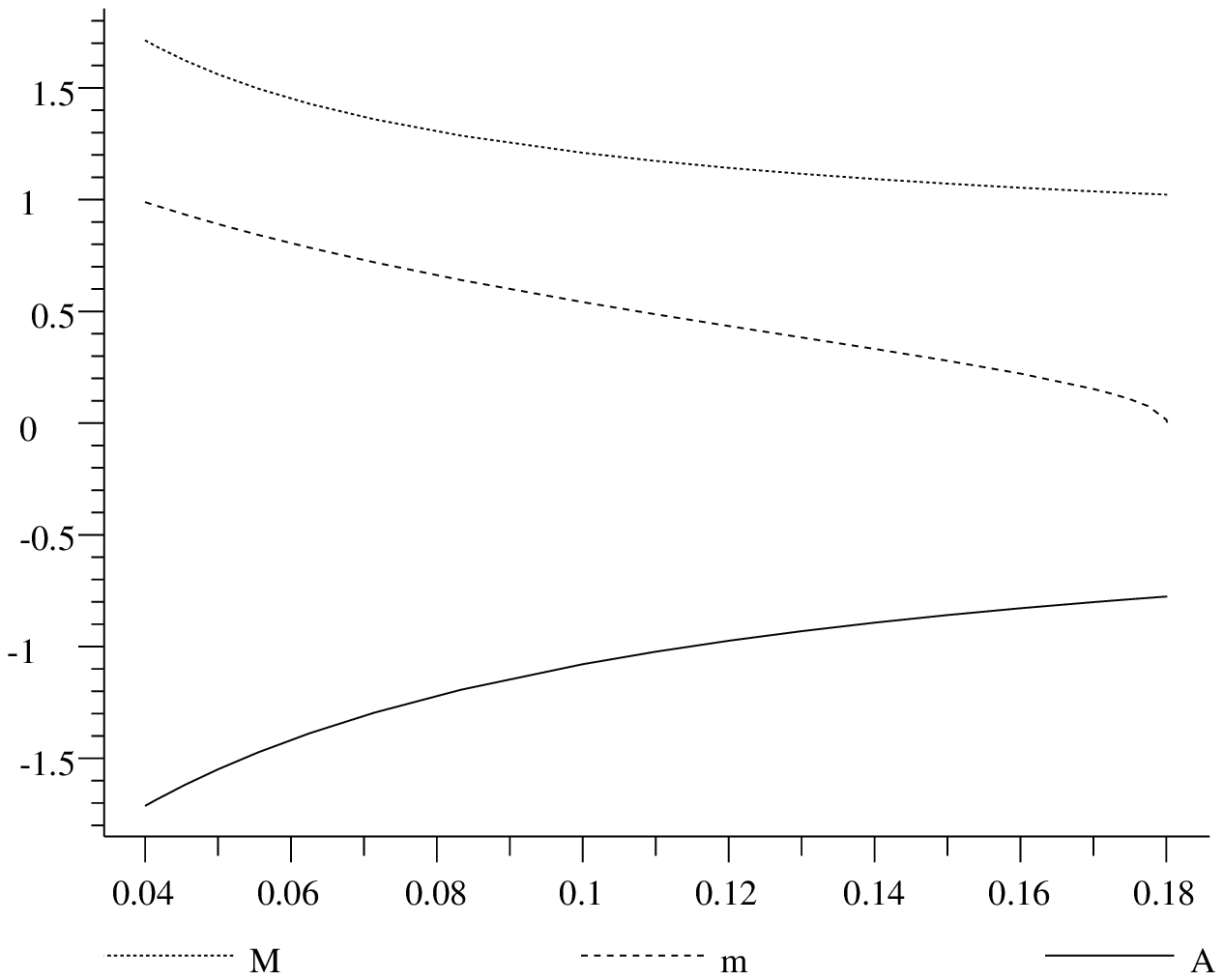,width=15cm}}
\bigskip
\caption[]{In M-theory model with $E_8$ as hidden sector group, 
the gaugino mass $M$, scalar mass $m$, and the trilinear coupling 
$A$ versus $\alpha_H$ in units of the gravitino mass.}
\label{diagrams}
\end{figure}

\begin{figure}
\centerline{\psfig{file=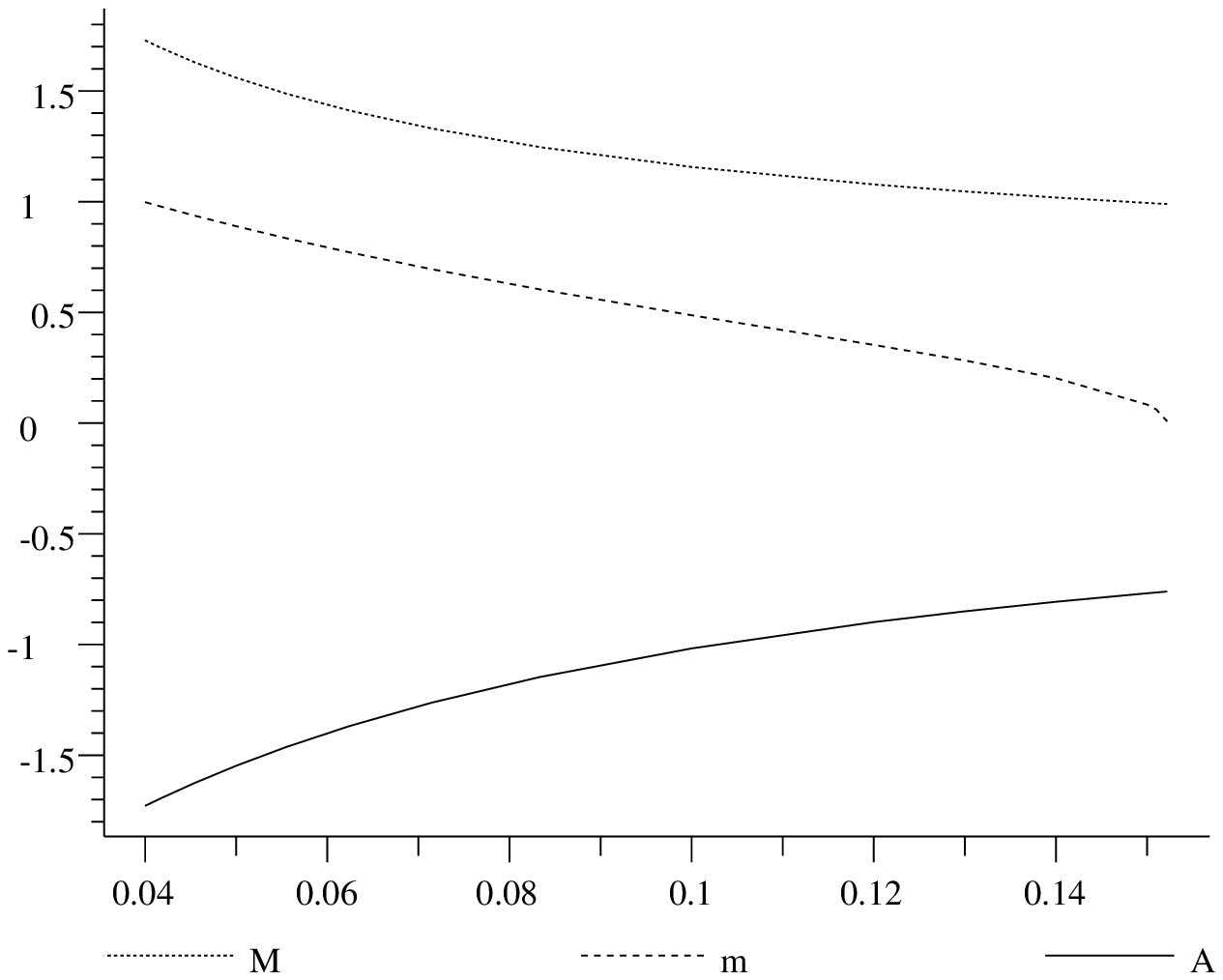,width=15cm}}
\bigskip
\caption[]{In M-theory model with $E_6$ as hidden sector group, 
the gaugino mass $M$, scalar mass $m$, and the trilinear coupling 
$A$ versus $\alpha_H$ in units of the gravitino mass.}
\label{diagrams}
\end{figure}

\end{document}